\documentclass[lettersize, journal]{IEEEtran}
\usepackage{setspace}
\usepackage{cite}
\usepackage{pgfplots}

\pgfplotsset{compat=1.10}

\usepgfplotslibrary{fillbetween}

\usepackage{bm}
\usepackage{lipsum}
\usepackage{makeidx}
\usepackage{enumerate}
\usepackage{color}
\usepackage{cite}
\usepackage{amsmath,amsthm}   
\usepackage{amssymb}
\usepackage{nomencl}
\usepackage{multirow}
\usepackage{graphicx}
\usepackage{epstopdf}
\usepackage{threeparttable}
\usepackage{multicol}
\DeclareGraphicsExtensions{.pdf,.jpeg,.png,.jpg,.emf,.eps}
\usepackage{calc}
\usepackage{here}
\usepackage{url}

\usepackage{upref}
\usepackage{comment}
\usepackage{times}
\usepackage{dsfont}
\usepackage{epic,eepic}
\usepackage{rawfonts}
\usepackage[T1]{fontenc}
\usepackage{latexsym}
\usepackage{amsfonts}

\usepackage{capitalgreekitalic}

\usepackage[font=small,labelfont=bf]{caption}
\usepackage[font=small,labelfont=bf]{subcaption}

\usepackage{xcolor}
\usepackage{tikz,pgfplots}
\usetikzlibrary{calc}
\usetikzlibrary{intersections}
\usetikzlibrary{arrows,shapes}
\usetikzlibrary{positioning}

\newtheorem{lemma}{Lemma}

\theoremstyle{definition}

\allowdisplaybreaks

\begin{document}
\title{\huge Spectral-Energy Efficiency Tradeoff of Nearly-Passive RIS in MIMO URLLC Downlink: Diagonal vs. Beyond Diagonal
}
\author{Mohammad Soleymani, \emph{Member, IEEE},  
Alessio Zappone, \emph{Fellow, IEEE}, 
Eduard Jorswieck, \emph{Fellow, IEEE},\\ Marco Di Renzo \emph{Fellow, IEEE},  
and Ignacio Santamaria, \emph{Senior Member, IEEE}, 
 \\ \thanks{ 
Mohammad Soleymani is with the Signal \& System Theory Group, Universit\"at Paderborn, Germany (email: \protect\url{mohammad.soleymani@uni-paderborn.de}).  

Alessio Zappone is with the Department of Electrical and Information
Engineering, University of Cassino and Southern Lazio, 03043 Cassino, Italy,
and also with CNIT, 43124 Parma, Italy (e-mail: \protect\url{alessio.zappone@unicas.it)}).

Eduard Jorswieck is with the Institute for Communications Technology, Technische Universit\"at Braunschweig, 38106 Braunschweig, Germany
(email: \protect\url{jorswieck@ifn.ing.tu-bs.de}).

M. Di Renzo is with Universit\'e Paris-Saclay, CNRS, CentraleSup\'elec, Laboratoire des Signaux et Syst\`emes, 3 Rue Joliot-Curie, 91192 Gif-sur-Yvette, France (email: \protect\url{marco.di-renzo@universite-paris-saclay.fr}), and with King's College London, Centre for Telecommunications Research -- Department of Engineering, WC2R 2LS London, United Kingdom (email: \protect\url{marco.di_renzo@kcl.ac.uk}).

Ignacio Santamaria is with the Department of Communications Engineering, Universidad de Cantabria, Spain (email: \protect\url{i.santamaria@unican.es}).

The work of A. Zappone has been supported by the EU through the HE-MSCA-TWIN6G project, grant agreement 101182794. The work of E. Jorswieck was supported by the Federal Ministry of Education and Research (BMBF, Germany) through the Program of Souver\"an. Digital. Vernetzt. joint Project 6G-RIC, under Grant 16KISK031, and by European Union's (EU's) Horizon Europe project 6G-SENSES under Grant 101139282. The work of M. Di Renzo was supported in part by the European Commission through the Horizon Europe project titled COVER under Grant 101086228, the Horizon Europe project titled UNITE under Grant 101129618, the Horizon Europe project titled INSTINCT under Grant 101139161, and the Horizon Europe project TWIN6G under grant agreement number 101182794, as well as by the Agence Nationale de la Recherche (ANR) through the France 2030 project titled ANR-PEPR Networks of the Future under Grant NF-PERSEUS 22-PEFT-004, and by the CHIST-ERA project titled PASSIONATE under Grants CHIST-ERA-22-WAI-04 and ANR-23-CHR4-0003-01. The work of I. Santamaria was funded by MCIN/AEI 10.13039/501100011033, under Grant PID2022-137099NB-C43 (MADDIE) and FEDER, UE, and by the Horizon Europe project 6G-SENSES under Grant 101139282. 
}}
\maketitle
\begin{abstract}
This paper investigates the spectral and energy efficiency (EE) tradeoff of nearly-passive (NP), both locally and globally NP (GNP), reconfigurable intelligent surface (RIS), considering diagonal and beyond-diagonal (BD) implementations in multi-user multiple-input multiple-output (MU-MIMO) broadcast channels designed for ultra-reliable low-latency communication (URLLC). We demonstrate that while all RIS architectures enhance the spectral efficiency, GNP BD-RIS achieves the highest gains. However, its EE is highly sensitive to the static circuit power consumption since BD-RIS has many more circuit elements than diagonal architectures. Furthermore, we demonstrate that the benefits of BD-RIS over diagonal RIS diminish as the number of data streams per user increases due to enhanced channel diversity in MIMO systems.
\end{abstract} 
\begin{IEEEkeywords}
Finite block length coding, low latency, max-min energy efficiency, max-min rate, MIMO systems,  reconfigurable intelligent surface, ultra-reliable communications.
\end{IEEEkeywords}%

\section{Introduction}\label{1} 
Energy efficiency (EE), spectral efficiency (SE), reliability, and latency have emerged as key performance indicators in modern wireless communication systems, with 6G networks expected to improve these metrics by at least an order of magnitude compared to 5G \cite{wang2023road, gong2022holographic}. A promising approach to significantly enhance these performance metrics is  the use of reconfigurable intelligent surfaces (RIS), which can operate in a nearly passive mode without requiring an external power supply \cite{wu2021intelligent, di2020smart}. 

There are two approaches to operate an RIS in nearly passive mode \cite{fotock2023energy, soleymani2024rate2}. In the first approach, each RIS element operates independently in a nearly passive manner, meaning the power of the reflected signal from each element does not exceed the power of the incident signal. This approach is referred to as the locally nearly-passive (LNP) mode. By contrast, the second approach allows some elements to amplify the incoming signal, provided that the total power of the reflected signal across all elements does not exceed the total received power. This more flexible strategy is referred to as globally nearly-passive (GNP). Note that unlike a hybrid RIS, which operates as an active device due to its net power gain \cite{huang2024hybrid}, the GNP RIS remains passive overall, as its total output power never exceeds the input power.

The scattering matrix of an RIS can be either diagonal or non-diagonal, depending on the underlying hardware architecture \cite{li2023beyond}. A diagonal scattering matrix corresponds to the conventional RIS design, where each element operates independently without requiring inter-element circuitry. In contrast, a non-diagonal scattering matrix, commonly called beyond-diagonal (BD), entails a more advanced design in which RIS elements are interconnected through dedicated circuits \cite{li2025tutorial}. This added flexibility enlarges the optimization space and enables more degrees of control in wave manipulation, often leading to higher SE \cite{santamaria2025rate, khan2025survey, maraqa2025beyond}. At the same time, BD-RIS introduces non-negligible implementation costs, mainly due to the higher static power consumption of the required circuitry \cite{soleymani2024energy}, necessitating careful analysis of its EE performance.

 The SE gains of GNP BD-RIS in multiple-input single-output (MISO) ultra-reliable low-latency communication (URLLC) systems were demonstrated in \cite{soleymani2024rate2}, while its EE performance was investigated in \cite{soleymani2024energy} for a MISO broadcast channel (BC). Both treatises, however, considered only single-antenna users and focused on a single performance indicator, namely SE in \cite{soleymani2024rate2} and EE in \cite{soleymani2024energy}. Moreover, they were limited to GNP BD-RIS, leaving locally passive (LNP) BD-RIS unaddressed. In this paper, we extend the analysis to multi-user (MU) multiple-input multiple-output (MIMO) URLLC BCs, jointly optimizing SE and EE. MU-MIMO systems introduce additional tradeoffs, as they exploit spatial diversity and enable multiple parallel data streams per user, which can significantly affect RIS performance. Furthermore, MU-MIMO requires more advanced analytical tools than MISO, particularly for efficient resource allocation. To the best of our knowledge, this is the first work to investigate BD-RIS, either LNP or GNP, in MU-MIMO URLLC systems.

  This work also differs fundamentally from the earlier studies on RIS-aided MU-MIMO URLLC systems in, e.g., \cite{soleymani2024optimization, soleymani2025framework}. These treatises focused exclusively on LNP diagonal RIS (D-RIS), whereas this paper studies three other RIS architectures, i.e., GNP D-RIS, LNP BD-RIS, and GNP BD-RIS, within a unified optimization framework. The feasibility sets of these new architectures are rigorously formulated to remain compatible with the models in \cite{soleymani2024optimization,soleymani2025framework}. Here, we extend the optimization framework to jointly capture the SE-EE tradeoff in BD-RIS-aided URLLC MU-MIMO systems, enabling a more comprehensive system-level analysis.

 The central message of this paper is built around a critical question, which is often overlooked in the literature: \textit{does the higher SE promised by more complex RIS architectures, such as GNP BD-RIS, justify the extra implementation cost in terms of energy consumption?} Our results show that the answer is not a simple yes or no. While BD-RIS consistently delivers superior SE compared to D-RIS, its EE performance is highly sensitive to the static power consumption of its circuit elements. Depending on the system parameters being considered, a simpler D-RIS may in fact provide better EE. This highlights that the tradeoff between added complexity and energy cost must be explicitly accounted for when evaluating the practicality of advanced RIS architectures.

The main contributions of this work are as follows. 
\begin{itemize}
    \item First, we show that all considered RIS architectures enhance the SE of the MU-MIMO URLLC BC, with the GNP BD-RIS architecture achieving the highest SE.
    
    \item Second, we analyze the EE performance and show that it strongly depends on the static power consumption of the BD-RIS circuitry. In particular, when the circuit elements are not sufficiently energy efficient, D-RIS can outperform BD-RIS, suggesting that switching some BD-RIS circuits off may be preferable in some cases. 
    
    \item Third, we demonstrate that the overall SE increases with the number of data streams per user, which holds for all scenarios, with or without RIS, and for all RIS architectures. However, the benefits of using RIS reduce as the number of data streams increases. In particular, the performance gap between GNP and LNP architectures, for both D-RIS and BD-RIS, also becomes smaller. This happens since MIMO systems with more data streams offer higher channel diversity, which reduces the impact of the RIS on system performance.

    \item  Finally, although the lower bounds used in the optimization framework are adapted from \cite{soleymani2024optimization}, the problem studied here, i.e., the SEE tradeoff of \textit{BD-RIS} in MU-MIMO URLLC systems, has not been addressed before. This enables us to directly examine the central question of whether the gains from added RIS complexities justify their costs.
\end{itemize}

{\em Notations}: Scalars/vectors/matrices are denoted by $s/{\bf s}/{\bf S}$. A zero-mean Gaussian signal ${\bf s}$ with covariance matrix ${\bf S}$ is represented as ${\bf s}\sim \mathcal{CN}({\bf 0}, {\bf S}  ) $. $\mathbb{E}\{ {\bf S}\}$, $\text{Tr}({\bf S})$, and $|{\bf S} |$ denote the mathematical expectation, trace, and determinant  of ${\bf S}$, respectively. ${\bf I}$ denotes the identity matrix.


\section{System model}\label{sec=ii} 
We consider an RIS-aided MIMO BC with a base station (BS), having $N_{BS}$ transmit antennas (TAs) and serving $K$ users with $N_u$ receive antennas (RAs) each. Also, an RIS with $N_{RIS}$ elements helps the BS. We assume that the data packets have a finite block length $n$. The transmit signal of the BS is 
${\bf x}=
\sum_k{\bf W}_k{\bf s}_{k},$
where ${\bf s}_{k}\sim \mathcal{CN}({\bf 0}, {\bf I}  ) $ is the message intended for user $k$, and the matrix ${\bf W}_k\in \mathbb{C}^{ N_{BS}\times N_{BS}}$ is the beamforming matrix, corresponding to ${\bf s}_{k}$. Thus, ${\bf x}$ is a zero-mean vector with the covariance matrix  
\begin{equation}\label{eq-cov}
{\bf C}\!=\!\mathbb{E}\{{\bf x}{\bf x}^H\}\!
=\sum_k\!{\bf W}_{k}{\bf W}_{k}^H\!\in\! \mathbb{C}^{ N_{BS}\times N_{BS}}\!.
\end{equation}
\subsection{RIS Model}\label{sec-ris}
We employ the RIS model in \cite{soleymani2022improper,pan2020multicell} for the MIMO BC. Therefore, the channel between the BS and user $k$ is
$\mathbf{H}_{k}\left({\bf \Phi}\right)= 
{\mathbf{G}_{k}{\bf \Phi}\mathbf{G}}+
{\mathbf{F}_{k}}$
,
where ${\bf F}_{k}$ is the channel between the BS and user ${k}$, ${\bf G}_{k}$ is the channel between the RIS and user $k$, ${\bf G}$ is the channel between the BS and the RIS, ${\bf \Phi}$  is the scattering matrix of the RIS. We consider the following RIS architectures. 

\subsubsection{Locally Nearly-Passive Diagonal RIS}
 In this architecture, the non-diagonal elements of ${\bf \Phi}$ are zero. Moreover, the absolute value of each diagonal element of ${\bf \Phi}$ is not greater than one ($|\phi_{mm}| \leq 1$). Hence, the feasible values of ${\bf \Phi}$ belong to the set
    $\mathcal{E}_{LPD}=\{\phi_{mn}: |\phi_{mm}| \leq 1,\phi_{mn} = 0,\forall m\neq n \}$.

\subsubsection{Locally Nearly-Passive Beyond Diagonal RIS}
This architecture realizes non-diagonal ${\bf \Phi}$, satisfying ${\bf \Phi}{\bf \Phi}^H\preceq {\bf I}$ and ${\bf \Phi}={\bf \Phi}^T$ \cite{li2022beyond}. We can rewrite ${\bf \Phi}{\bf \Phi}^H \preceq{\bf I}$ as 
    $\left[\begin{array}{cc}
         {\bf I}&{\bf \Phi}  \\
         {\bf \Phi}^H& {\bf I}
    \end{array} \right]\succeq {\bf 0}$
to easier implement this convex constraint in numerical solvers. Therefore, the feasible set of ${\bf \Phi}$ in this architecture is
    $\mathcal{E}_{LPBD}\!=\!\{{\bf \Phi}: {\bf \Phi}={\bf \Phi}^T,{\bf \Phi}{\bf \Phi}^H \preceq{\bf I} \}$.

\subsubsection{Globally Nearly-Passive Diagonal RIS}
Following the definition in \cite{fotock2023energy} and \cite{soleymani2024energy}, ${\bf \Phi}$ should fulfill the constraint
\begin{equation}\label{(4)}
    p_{out}-p_{in}=\text{Tr}
    \left( 
    {\bf G}{\bf C}{\bf G}^H
    ({\bf \Phi}^H{\bf \Phi}-{\bf I}_M)\right)\leq 0,
\end{equation}
where $ p_{out}$ and $p_{in}$ are the output and input power to the RIS, respectively, and ${\bf C}$ is defined in \eqref{eq-cov}. 
Hence, the feasible set of ${\bf \Phi}$ is:
\begin{equation}
       \scalebox{0.9}{\ensuremath{\mathcal{E}_{D}\!=\!\{\!{\bf \Phi}\!:\!\text{Tr}
    \left( 
    {\bf G}{\bf C}{\bf G}^H
    ({\bf \Phi}^H{\bf \Phi}-{\bf I}_M)\right)\leq 0,
    \phi_{mn} = 0,\forall m\neq n\}}}.\!
\end{equation}

\subsubsection{Globally Nearly-Passive Beyond Diagonal RIS}
In this case, ${\bf \Phi}$ can be a non-diagonal symmetric matrix, but it needs to fulfill \eqref{(4)}. Thus, we have
\begin{equation}\label{(6)}
    \mathcal{E}_{BD}\!=\!\{{\bf \Phi}: {\bf \Phi}={\bf \Phi}^T,\text{Tr}
    \left( 
    {\bf G} {\bf C} {\bf G}^H
    ({\bf \Phi}^H{\bf \Phi}-{\bf I}_M)\right)\leq 0 \}.
\end{equation}

\subsection{Rate and EE Expressions} 
The received signal of user $k$ is
${\bf y}_{k}={\bf H}_{k}\sum_i{\bf W}_i{\bf s}_i+{\bf n}_{k}$,
where ${\bf n}_{k}\sim \mathcal{CN}({\bf 0}, \sigma^2 {\bf I} )$ is the additive noise at the receiver of user ${k}$, and $\sigma^2$ is the noise variance at each receive antenna.
Upon utilizing the normal approximation (NA), the finite block length (FBL) rate of user $k$ is \cite{polyanskiy2010, soleymani2024optimization}
\begin{multline}
r_k\left(\{{\bf W} \},{\bf \Phi} \right)=
\log \left|{\bf I} +\left(\sigma^2{\bf I}+\sum_{i\neq k}{\bf Z}_{ki}\right)^{-1}{\bf Z}_{kk} \right|
\\-\frac{Q^{-1}(\epsilon)}{\sqrt{n}}\sqrt{2\text{Tr}\left({\bf Z}_{kk}\left(\sigma^2{\bf I}+\sum_{i}{\bf Z}_{ki}\right)^{-1}\right)},
  \label{1-multi}
\end{multline}
where $\epsilon$  is the maximum tolerable bit error rate for each message, $n$ is the codeword length of each message, and ${\bf Z}_{ki}={\bf H}_{k}{\bf W}_{i}{\bf W}_{i}^H{\bf H}_{k}^H$ for all $i,k$.

 The EE of user ${k}$ is 
     $e_{k}\left(\{{\bf W} \},{\bf \Phi} \right)=\frac{r_{k}\left(\{{\bf W} \},{\bf \Phi} \right)}{P_c+\eta\text{Tr}\left({\bf W}_{k}{\bf W}_{k}^H\right)
     }$,
 where 
$\eta^{-1}$ is the power efficiency of the BS, and $P_c$ is the static power consumption is given by \cite[Eq. (11)]{soleymani2024energy}
\begin{equation}\label{(p-c)}
   P_c=\frac{P_{BS}+P_{RIS}}{K}+P_{UE}=P_t+\frac{P_{RIS}}{K},
\end{equation}
where $P_{BS}$ and $P_{UE}$ are defined in  \cite[Eq. (11)]{soleymani2024energy}. 
Moreover, $P_{RIS}$ is the static power to operate the RIS. 
For a nearly passive D-RIS architecture, the static power consumption is given by
\begin{equation}\label{(12)}
    P_{RIS}=P^{D}_{RIS,0}+N_{RIS}P^{D}_{RIS,n},
\end{equation}
where $P^{D}_{RIS,0}$ denotes the static power of the diagonal architecture,
and $P^{D}_{RIS,n}$ is the static power of each RIS element, and $N_{RIS}$
is the number of RIS elements. Moreover, the static power consumption of a GNP BD-RIS is
\begin{equation}\label{(13)}
    P_{RIS}=P^{BD}_{RIS,0}+N_cP^{BD}_{RIS,n},
\end{equation}
where $P^{BD}_{RIS,0}$ is the static power of the BD architecture, $P^{BD}_{RIS,n}$ is the static power of each circuit element, and $N_c$ represents the total number of circuit elements of GNP BD-RIS. In this paper, we consider a fully-connected GNP BD-RIS architecture, for which $N_c = N_{RIS}(N_{RIS}-1)/2$.
\subsection{Problem Statement} 
 {A metric to evaluate the SEE tradeoff is \cite{soleymani2025framework}
\begin{equation}\label{eq-see}
   \zeta_k=\alpha_k r_k+(1-\alpha_k) e_k=\alpha_k r_k+ \frac{(1-\alpha_k) r_k}{P_c+\eta \text{Tr}({\bf W}_k{\bf W}_k^H) },
\end{equation}
where $0\leq\alpha_k\leq 1$ and $1-\alpha_k$ are the weights of the SE and EE for user $k$, respectively. These weights correspond to the priority of the SE or EE for each individual user.}
Maximizing the minimum of $\zeta_k$ for all $k$ yield
    \begin{align}\label{eq-opt}
       \scalebox{1}{\ensuremath{\!\! \underset{\{{\bf W}\},{\bf \Phi}\in\mathcal{E} }{\max} \underset{k}{\min}\left\{
        \alpha_k r_k
        \!+\!\frac{(1-\alpha_k) r_k}{P_c+\eta \text{Tr}({\bf W}_k{\bf W}_k^H) }
        \right\}\ 
        \text{s.t.}\ 
         %
        \text{Tr}({\bf C})\!\leq\! P\!,}}\!\!
    \end{align}
where $P$ is the power budget of the BS, and $\mathcal{E}$ is the set of all feasible ${\bf \Phi}$, which can be $\mathcal{E}_{LPD}$, $\mathcal{E}_{LPBD}$, $\mathcal{E}_{D}$, or $\mathcal{E}_{BD}$, depending on the RIS architecture. Note that $\mathcal{E}$ is a convex set in either case. The objective function of \eqref{eq-opt} has a fractional structure, and, therefore, \eqref{eq-opt} is a fractional matrix programming (FMP) optimization problem.

\section{Proposed solution} 
 {To solve \eqref{eq-opt}, we employ alternating optimization (AO), the FMP solver proposed in \cite{soleymani2025framework}, and the lower bounds in \cite{soleymani2024optimization}. The proposed solution is iterative, starting with a feasible initial point $\{{\bf W}^{(1)} \}$ and ${\bf \Phi}^{(1)}$. In iteration $l$, we first update $\{{\bf W}   \}$, keeping ${\bf \Phi}$ fixed at ${\bf \Phi}^{(l)}$. Then we alternate and update ${\bf \Phi}$ while $\{{\bf W}\}$ is fixed to $\{{\bf W}^{(l+1)}\}$. Each iteration  monotonically improves the objective function of \eqref{eq-opt}, and thus, the algorithm converges. Note that \cite{soleymani2024optimization, soleymani2025framework} do not optimize GNP D-RIS and BD-RIS architectures. In this treatise, the feasibility sets of these advanced RIS architectures are constructed in a manner consistent with \cite{soleymani2024optimization, soleymani2025framework}, requiring a nontrivial generalization of the prior frameworks.} In the following, we present the algorithms for updating $\{{\bf W}   \}$ and ${\bf \Phi}$.
\subsection{Updating $\{{\bf W}\}$ } 
When ${\bf \Phi}$ is fixed to ${\bf \Phi}^{(l)}$, \eqref{eq-opt} is simplified to
    \begin{align}\label{eq-opt-w}
        \max_{\{{\bf W}\} } & \min_k\!\left\{\!
        \alpha_k r_k
        \!+\!\frac{(1-\alpha_k) r_k}{P_c\!+\!\eta \text{Tr}({\bf W}_k{\bf W}_k^H) }\!
        \right\}\!\!
        &
        \text{s.t.}\,& 
         %
        \text{Tr}({\bf C})\!\leq\! P\!,
    \end{align}
which is an FMP problem in $\{{\bf W}\}$. We employ \cite[Lemma 2]{soleymani2025framework} to solve \eqref{eq-opt-w}. To this end, we require a concave lower bound for $r_k$, which is presented in the following lemma. 
\begin{lemma}[\!\cite{soleymani2024optimization}]  
	\label{lem:1}
	All feasible  beamforming matrices satisfy
    \begin{multline*}
        r_{k}(\{{\bf W}\})\! \geq\! \tilde{r}_{k}(\{{\bf W}\})\!=\! a_{k}
+2\sum_{i}\!\mathfrak{R}\!\left\{\!\text{{\em Tr}}\!\left(
{\bf A}_{kj}{\bf W}_{i}^H
\mathbf{H}_{k}^{(l)^H}\right)\!\!\right\}
\\-
\text{{\em Tr}}\big(
{\bf B}_{k}\big(\sigma^2{\bf I}
+\sum_j\mathbf{H}_{k}^{(l)}{\bf W}_{i}{\bf W}_{i}^H\mathbf{H}_{k}^{(l)^H}\big)
\big),
    \end{multline*}
     {where $\mathbf{H}_{k}^{(l)}= 
{\mathbf{G}_{k}{\bf \Phi}^{(l)}\mathbf{G}}+
{\mathbf{F}_{k}}$, and the coefficients $a_k$, ${\bf A}_{kj}$, and ${\bf B}_k$ for all $k,j$ are defined as in \cite[Lemma 5]{soleymani2024optimization}.}
\end{lemma}
 Upon using Lemma \ref{lem:1} and \cite[Lemma 2]{soleymani2025framework}, we can transform \eqref{eq-opt-w} to the following convex optimization problem  
\begin{subequations}\label{eq-opt-w3}
    \begin{align}
        \max_{\{{\bf W}\},{\bf u} } & \scalebox{0.96}{\ensuremath{\min_k\!
        \left\{\!
        \alpha_k \tilde{r}_k(\{{\bf W}\})
        \!+\!(1-\alpha_k)\!
        \left(\!2\beta_k^{(l)} u_k\!-\!\beta_k^{(l)^2}p_k \right)\!
        \right\}\!\!}}
        \\
         \text{s.t.}\,\,& 
        \text{Tr}({\bf C})\leq P,\,\,\,\tilde{r}_k(\{{\bf W}\})-u_k^2\geq 0,\,\,\, \forall k,
    \end{align}
\end{subequations}
where $\beta_k^{(l)}=\frac{\sqrt{r_k({\bf W}^{(l)},{\bf \Phi}^{(l)})}}{P_c+\eta \text{Tr}\left({\bf W}_k^{(l)}{\bf W}_k^{(l)^H}\right) }$ is a coefficient, $p_k=P_c+\eta \text{Tr}\left({\bf W}_k{\bf W}_k^H\right)$, and ${\bf u}=[u_1,u_2,\cdots,u_K] $ is a set of auxiliary optimization variables. The solution of \eqref{eq-opt-w3} is $\{{\bf W}^{(l+1)} \}$.

\subsection{Updating ${\bf \Phi}$}
When the beamforming matrices are fixed, \eqref{eq-opt} is equivalent to maximizing the minimum weighted rates in ${\bf \Phi}$ as 
    \begin{align}\label{eq-opt-phi}
        \max_{{\bf \Phi}\in\mathcal{E} } & \min_k\!\left\{\!\!
        \!\left(\!\!\alpha_k\! 
        +\!\frac{(1-\alpha_k)}{P_c\!+\!\eta \text{Tr}\!\left(\!{\bf W}_k^{(l+1)}{\bf W}_k^{(l+1)^H}\!\right) }
        \right)\!r_k({\bf \Phi})\!
        \!\right\}\!,\!\!
         %
    \end{align}
where $\alpha_k 
        +\frac{(1-\alpha_k)}{P_c+\eta \text{Tr}\left({\bf W}_k^{(l+1)}{\bf W}_k^{(l+1)^H}\right) }$ is a constant coefficient. 
The set $\mathcal{E}$ is convex for all the architectures considered in Section \ref{sec-ris}. However, \eqref{eq-opt-phi} is non-convex since $r_k$ is non-concave in ${\bf \Phi}$. To solve \eqref{eq-opt-phi}, we first obtain a concave lower bound for $r_k$  as in the following lemma.
\begin{lemma}[\!\cite{soleymani2024optimization}]
	\label{lem:2}
	All the feasible channels for ${\bf \Phi}\in\mathcal{E}$ obey
    \begin{multline*}
        r_{k}({\bf \Phi})\! \geq\! \hat{r}_{k}({\bf \Phi})\!=\! a_{k}\!
+\!2\sum_{i}\mathfrak{R}\!\left\{\!\text{{\em Tr}}\!\left(\!
{\bf A}_{kj}{\bf W}_{i}^{(l+1)^H}
\mathbf{H}_{k}^H({\bf \Phi})\!\right)\!\right\}
\\-
\text{{\em Tr}}\left(
{\bf B}_{k}\left(\sigma^2{\bf I}
+\sum_j\mathbf{H}_{k}({\bf \Phi}){\bf W}_{i}^{(l+1)}{\bf W}_{i}^{(l+1)^H}\mathbf{H}_{k}^H({\bf \Phi})\right)
\right),
    \end{multline*}
    where all coefficients are defined as in Lemma \ref{lem:1}. 
\end{lemma}
Upon utilizing $\hat{r}_k$, we can update  ${\bf \Phi}$ by solving the convex problem
    \begin{align}\label{eq-opt-phi2}
        \max_{{\bf \Phi}\in\mathcal{E} } & \min_k\!\left\{\!\!\!
        \left(\!\!\alpha_k \!
        +\!\frac{(1-\alpha_k)}{P_c+\eta \text{Tr}\left({\bf W}_k^{(l+1)}{\bf W}_k^{(l+1)^H}\!\!\right) }
        \right)\!\hat{r}_k({\bf \Phi})
        \!\!\right\}\!.\!\!
         %
    \end{align}

\subsection{Computational Complexity Analysis} 
In this subsection, we calculate an approximate upper bound for the number of multiplications imposed by running our algorithms. Each iteration of our proposed framework consists of two steps, as summarized in Algorithm I. In the first step, 
 we update $\{{\bf W}\} $ by solving \eqref{eq-opt-w3}. An approximate number of multiplications for solving \eqref{eq-opt-w3} can be obtained similar to \cite[Sec. VI.B]{soleymani2025framework}, which is $\mathcal{O}\left[N_{BS}^2K^2\sqrt{2K+1}(2N_{BS}+N_{u})\right]$.

Now, we derive an approximation for the number of multiplications needed to update $\{{\bf \Phi}\}$, i.e., the number of multiplications needed for solving \eqref{eq-opt-phi2}. To solve each Newton iteration, we have to compute $K$ surrogate functions for the rates, $\hat{r}_{k}$, as well as $K$ equivalent channels, ${\bf H}_k({\bf \Phi}) $. To compute each channel, we need approximately $N_{u}N_{BS}N_{RIS}$ multiplications for D-RIS and $N_{u}N_{BS}N_{RIS}^2$ multiplications for BD-RIS. Moreover, the computational complexity of calculating $\hat{r}_{k}$  is approximately $\mathcal{O}\left[ K N_{BS}^2(2N_{BS}+N_{u})\right]$. Finally, the computational complexity of updating ${\bf \Phi}$ can be approximated as $\mathcal{O}[K( K N_{BS}^2(2N_{BS}+N_{u})+N_{u}N_{BS}N_{RIS}^2)]$ for BD-RIS and $\mathcal{O}[K( K N_{BS}^2(2N_{BS}+N_{u})+N_{u}N_{BS}N_{RIS})]$ for D-RIS. Note that the computational complexities of LNP and GNP are approximately of the same order, as the number of multiplications to compute \eqref{(4)} is significantly lower than those of calculating $K$ channel matrices and $K$ surrogate functions for the rates.

\doublespacing 
\begin{table}[htb]
\begin{tabular}{l}
\hline 
 \textbf{Algorithm I}: Summary of the proposed solution.  \\
\hline
\hspace{0.2cm}{\textbf{Initialization}}\\
\hspace{0.2cm}Set $\delta$,  $l=1$,  $\{\mathbf{W}\}=\{\mathbf{W}^{(1)}\}$, ${\bf \Phi}={\bf \Phi}^{(1)}$ \\
\hline 
\hspace{0.2cm}
\textbf{While} $\left(\underset{\forall k}{\min}\,\zeta^{(l+1)}_{k}-\underset{\forall k}{\min}\,\zeta^{(l)}_{k}\right)/\underset{\forall k}{\min}\,\zeta^{(l)}_{k}\geq\delta$\\ 
\hspace{.6cm}{Derive $\tilde{r}_{k}$, 
employing Lemma \ref{lem:1}}\\ 
\hspace{.6cm}{Compute $\{{\bf W}^{(l+1)} \}$ by solving \eqref{eq-opt-w3}}\\
\hspace{.6cm}{Obtain $\hat{r}_{k}$, 
employing Lemma \ref{lem:2}}\\ 
\hspace{.6cm}{Find ${\bf \Phi}^{(l+1)} $ by solving \eqref{eq-opt-phi2}}\\
\hspace{.6cm}$l=l+1$\\
\hspace{0.2cm}\textbf{End (While)}\\
\hspace{0.2cm}{{\bf Return} $\{\mathbf{W}^{(\star)}\}$ and ${\bf \Phi}^{(\star)}$}\\
\hline 
\end{tabular}  
\end{table}
\singlespacing\normalsize
\section{Numerical results}\label{sec-iv}
Monte Carlo simulations are leveraged to evaluate the performance of the nearly passive RIS architectures studied in this paper. The channel models, simulation scenario, and parameters are based on \cite[Appendix G]{soleymani2025framework}. Specifically, the BS-to-RIS channel is modeled using a Rician distribution with a Rician factor equal to three. By contrast, the RIS-to-user channels are assumed to be non-line-of-sight (NLOS), following a Rayleigh fading model. The heights of the BS and RIS are set to 25 meters, while the height of users is 1.5 meters. The BS is positioned at the origin, with the RIS is located 130 meters away. Users are randomly distributed within a square region of 20 meters per side, centered at the RIS position. For a baseline comparison, a configuration without RIS (labeled as No-RIS) is considered, where the BS-to-user links are Rayleigh-distributed due to NLOS propagation. Additionally, we include a scenario where LNP D-RIS reflection coefficients are assigned randomly (RIS-Rand). Moreover, we assume that the weights for the SE and EE of each user is the same, i.e., $\alpha_k=\alpha$ for all $ k$.

\begin{figure}[t]
    \centering
    \begin{subfigure}[t]{0.48\textwidth}
        \centering
           \includegraphics[width=\textwidth]{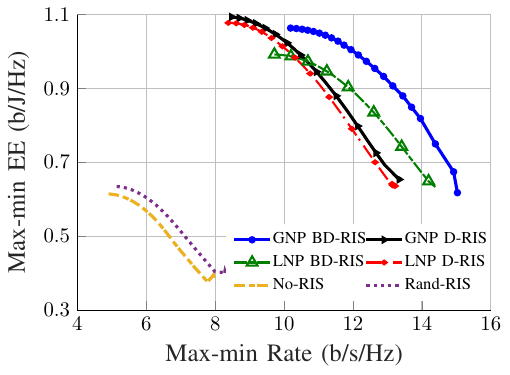}
        \caption{$P_{RIS,n}=20$ mW.}
    \end{subfigure}
\begin{subfigure}[t]{0.48\textwidth}
        \centering
       \includegraphics[width=\textwidth]{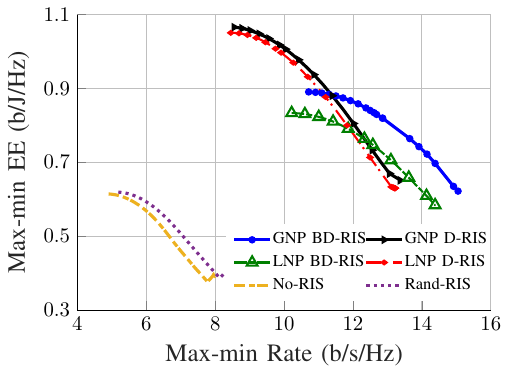}
        \caption{$P_{RIS,n}=40$ mW.}
    \end{subfigure}%
    \caption{Max-min EE versus max-min rate ( $P=10$dB, $N_{{BS}}=N_{u}=4$, $K=2$,   and $N_{{RIS}}=20$).}
	\label{Fig-pareto}   
\end{figure}

Fig. \ref{Fig-pareto} illustrates the max-min SEE tradeoff, where the x-axis represents the max-min rate and the y-axis shows the corresponding max-min EE  for $\alpha_k=\alpha,\ \forall k$. The figure captures all the achievable SE–EE operating points. As observed, RIS can substantially improve both SE and EE when optimized. Moreover, GNP D-RIS outperforms LNP D-RIS, though the performance gap remains relatively small. Both LNP and GNP BD-RIS offer higher SE, but their EE may be less than the EE of the D-RIS architectures, especially when the static power is high. In particular, for large values of $P_{RIS,n}$, we observe an intersection between the Pareto boundaries of GNP BD-RIS and D-RIS, indicating that the circuit design of BD-RIS should be carefully optimized when the EE is the primary concern.

\begin{figure}[t]
    \centering
           \includegraphics[width=.44\textwidth]{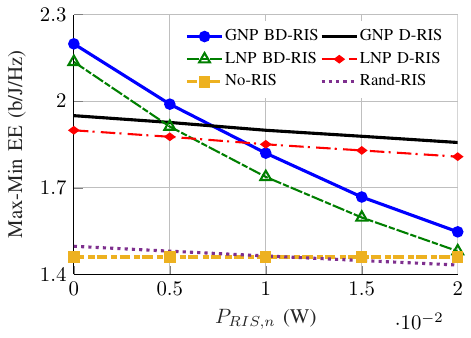}
    \caption{Average max-min EE versus 
 $P_{RIS,n}$ ($P_t=2.5$ W, $N_{BS}=N_u=5$, $K=2$, $N_{RIS}=20$ and $P=10$ dB).}
	\label{Fig-ee}  
\end{figure}
Fig. \ref{Fig-ee} depicts the average max-min EE versus $P_{RIS,n}$, showing that all RIS architectures enhance the EE when carefully optimized. The EE of all RIS architecture reduces when $P_{RIS,n}$ increases. However, this performance degradation is mild for diagonal architectures because of fewer circuit elements. Hence, both LNP and GNP D-RISs maintain a significant EE gain even when $P_{RIS,n}$ is relatively high. By contrast, the EE of BD-RIS significantly drops with $P_{RIS,n}$, and it may even degrade the overall EE when the static power is high due to its large number of circuit elements.

\begin{figure}[t]
    \centering
           \includegraphics[width=.44\textwidth]{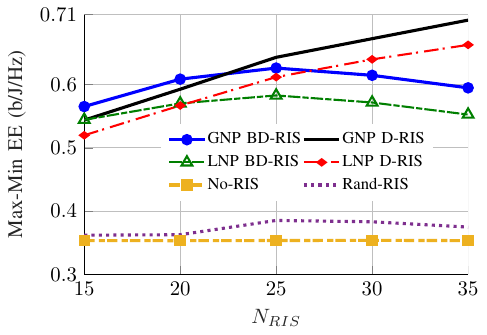}
    \caption{Average max-min EE versus 
 $N_{RIS}$ ($P_t=5$ W, $P_{RIS,n}=10$ mW, $N_{BS}=N_u=2$, $K=2$ and $P=10$ dB).} 
	\label{Fig-ee2} 
\end{figure}
 {Fig. \ref{Fig-ee2} illustrates the average max-min EE versus $N_{RIS}$. All RIS architectures enhance EE compared to No-RIS, and their performance initially improves as $N_{RIS}$ increases. When the number of RIS elements is small, BD-RIS architectures achieve higher EE than D-RIS due to their richer configuration flexibility. However, as $N_{RIS}$ grows, the static power consumption of BD-RIS, which scales quadratically  with $N_{RIS}$ because of the inter-element circuitry, becomes dominant. Consequently, the EE of BD-RIS saturates and may even decline, while the EE of D-RIS continues to increase almost linearly with $N_{RIS}$. This figure highlights the tradeoff between performance and complexity: although both D-RIS and BD-RIS achieve higher SE with more reflecting elements, the rapidly growing implementation complexity and power consumption of BD-RIS may ultimately outweigh its SE gains.}

\begin{figure}[t]
    \centering
    \begin{subfigure}[t]{0.4\textwidth}
        \centering
           \includegraphics[width=\textwidth]{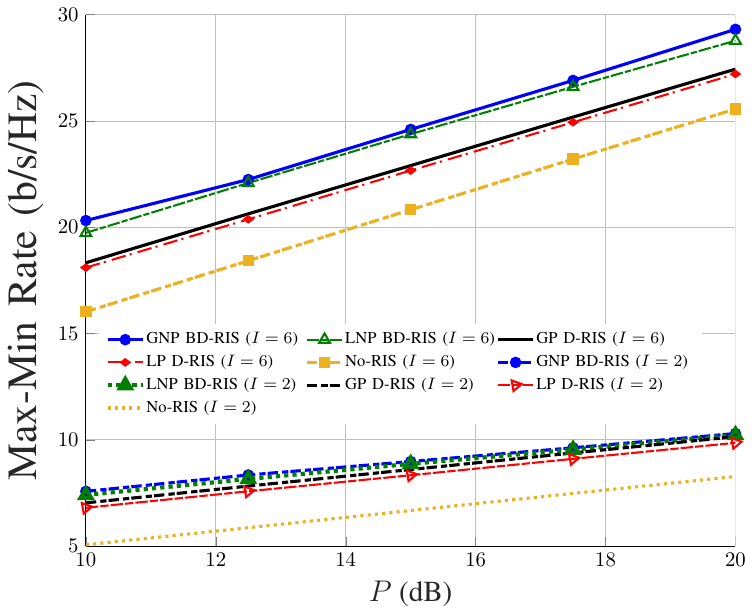}
        \caption{Average max-min rate.}
    \end{subfigure}
\begin{subfigure}[t]{0.4\textwidth}
        \centering
       \includegraphics[width=\textwidth]{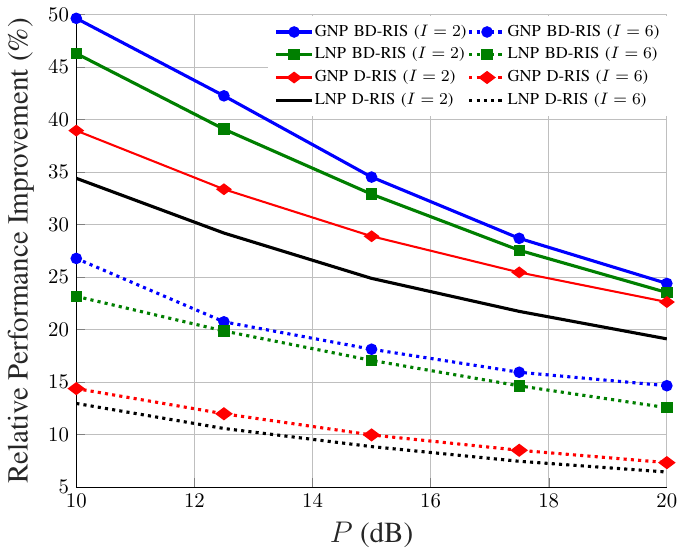}
        \caption{Gains of RIS architectures.}
    \end{subfigure}%
    \caption{SE versus $P$  ($I=N_{BS}=N_u$, $K=2$,  and $N_{RIS}=20$).}
	\label{Fig-se-ben}   
\end{figure}

Fig. \ref{Fig-se-ben} shows the SE performance of RIS architectures versus the BS power as a function of the maximum number of data streams per users $I=\min(N_{BS},N_u)$. As observed, GNP BD-RIS outperforms the other RIS architectures; however, the performance gap between LNP and GNP BD-RIS is not substantial. Moreover, the benefits of all RIS architectures diminish as $I$ increases. This is because the spatial diversity provided by MIMO reduces the relative impact of the RIS configuration. Additionally, the performance gain obtained by GNP BD-RIS is more prominent at lower SNR, while it becomes marginal compared to the gain of LNP BD-RIS at high-SNR.


\section{Summary and Conclusions}\label{sec-v}
This paper studied the SEE tradeoff of nearly passive RIS architectures in a MU-MIMO URLLC BC, showing that GNP BD-RIS enhances the max-min rate and EE compared to systems operating without RIS. GNP BD-RIS is more general than other nearly passive architectures; hence, an optimal GNP BD-RIS never performs worse than the other RIS architectures in terms of SE. However, BD-RIS (both LNP and GNP) may be less energy efficient than its diagonal counterpart, as BD-RIS has a much higher number of circuit elements, yielding a higher static power consumption. Hence, if the EE is the primary concern, the number of circuit elements enabling BD-RIS needs to be carefully optimized to balance performance and power consumption. Future work may consider alternative BD-RIS architectures with fewer interconnections such as band-connected and stem-connected designs to further improve EE \cite{wu2025beyond}.


\bibliographystyle{IEEEtran}
\bibliography{ref}
\end{document}